\documentclass[%
 showpacs,
 reprint,
 groupedaddress,
 showpacs,
 amsmath,amssymb,
 aps,
prb,
]{revtex4-1}

\usepackage[pdftex]{graphicx, color}
\usepackage{multirow,booktabs}
\usepackage{bm}%
\usepackage{braket}
\usepackage{siunitx}
\usepackage{hyperref}
\usepackage{ulem}

\DeclareMathOperator{\Tr}{Tr}
\DeclareMathOperator{\tr}{tr}

\newcommand{\Eq}[1]{Eq.~\eqref{#1}}

\newcommand{\Fig}[1]{Fig.~\ref{#1}}
\renewcommand{\Ref}[1]{Ref.~\cite{#1}}
\newcommand{\Sec}[1]{Sec.~\ref{#1}}
\newcommand{\Cite}[1]{~\cite{#1}}

\begin{document}

\title{Thermodynamic approach to electric quadrupole moments}

\author{Akito Daido}
\email[]{daido@scphys.kyoto-u.ac.jp}
\affiliation{%
 Department of Physics, Graduate School of Science, Kyoto University, Kyoto 606-8502, Japan
}%

\author{Atsuo Shitade}
\affiliation{%
 Institute for Molecular Science, Aichi 444-8585, Japan
}%

\author{Youichi Yanase}
\affiliation{%
 Department of Physics, Graduate School of Science, Kyoto University, Kyoto 606-8502, Japan
}%

\date{\today}

\begin{abstract}
  Higher-rank electric/magnetic multipole moments are attracting attention these days as candidate order parameters for exotic material phases.
  However, quantum-mechanical formulation of those multipole moments is still an ongoing issue.
  In this paper, we propose a thermodynamic definition of electric quadrupole moments as a measure of symmetry breaking, following previous studies of orbital magnetic dipole moments and magnetic quadrupole moments.
  The obtained formulas are illustrated with a model of orbital-ordered nematic phases of iron-based superconductors.
\end{abstract}

\maketitle

\section{Introduction}
Recent years have witnessed a variety of exotic material phases which were not recognized several decades ago.
Among them, intriguing phase are those beyond description by simple order parameters. For example, electronic nematic phases observed in iron-based superconductors\Cite{chuang2010,chu2010,kasahara2012} cannot be described by the electric polarization or the magnetization owing to the inversion and time-reversal symmetries.
Some materials including URu$_2$Si$_2$ and Sr$_2$IrO$_4$ show mysterious phases even called ``hidden order"\Cite{palstra1985,MydoshRMP2011,shibauchi2014,Zhao2016,Jeong2017,murayama2020}, which are literally beyond the scope of familiar order parameters.
Proper characterization and understanding of these novel phases is one of the central issues of modern condensed matter physics.

There is renewed interest in higher-rank electric and magnetic multipole moments\Cite{Jackson}, because of their potential ability to describe the novel material phases.
Each electric/magnetic multipole moment is categorized into either one of the irreducible representations of the point group,
giving us good labeling of the symmetry of ordered phases\Cite{watanabe2018,hayami2018}.
A familiar example is the electric polarization, which is widely accepted as the order parameter of the inversion symmetry breaking.
Electric quadrupole moments (EQMs) may play the same role for fourfold rotation ($C_4$) symmetry breaking, as actually have been discussed in $f$ electron systems.
The observed $C_4$ symmetry breaking in CeAg, UNiSn, PrTi$_2$Al$_{20}$, and so on, is consistent with the ferroic alignment of atomic-scale EQMs of $f$ electrons~\cite{morin1988,akazawa1996,onimaru2016}.
As for magnetic multipole moments, magnetic quadrupole moments (MQMs) are among the hottest topics in recent research works, owing to their close relationship with the magnetoelctric effect and multiferroics~\cite{Spaldin2008}.
Magnetic octupole and dotriacontapole, as well as electric quadrupole and hexadecapole moments are proposed as candidate order parameters for the hidden order in URu$_2$Si$_2$~\cite{MydoshRMP2011,shibauchi2014}, while anapole moment is implied as a potential hidden order parameter in Sr$_2$IrO$_4$~\cite{Zhao2016,Jeong2017,murayama2020}.
Some of these compounds undergo superconductivity, and therefore, a relation between the multipole order and superconductivity is also a topic of interest~\cite{sumita2020}.

We note that an attention must be paid to the meaning of higher-rank multipole moments in the previous studies:
They refer to only the symmetry and/or microscopic degrees of freedom, rather than those possessed by the whole crystal.
Actually, quantum-mechanical formulation of higher-rank multipole moments is still an ongoing issue, due the unbounded nature of the position operator.
In contrast to the electric polarization\Cite{KingSmith1993,Vanderbilt1993,Vanderbilt2018} and the orbital magnetization\Cite{Xiao2005,Thonhauser2005,Ceresoli2006,Souza2008,Shi2007}, which were established in early days, it has been revealed only recently that EQMs are well formulated as topological invariants of higher-order topological crystalline insulators\Cite{Benalcazar2017a,Benalcazar2017b,Song2017,Ezawa2018,Franca2018,Benalcazar2019,Schindler2019,Watanabe2020,Hirayama2020,Imhof2018,Serra-Garcia2018,Peterson2018,%
He2020,Peterson2020}.
The obtained formulas for EQMs are, however, valid only in the presence of crystalline symmetries which force EQMs to vanish in a classical sense (e.g. $C_4$)\Cite{Benalcazar2017b}.
Thus, the formula is not applicable to EQMs emerging as a result of symmetry breaking.
Although an extension of Refs.~\cite{Benalcazar2017a}~and~\cite{Benalcazar2017b} to systems without those crystalline symmetries has been discussed by introducing a generalized twist operator\Cite{Kang2019,Wheeler2019}, there exists some discrepancy due to the mismatch with periodic boundary conditions\Cite{Wheeler2019,Ono2019}.
Thus, there is no systematic way to quantify EQMs of $C_4$-breaking materials so far.

As noted in electromagnetism\Cite{Jackson} and also in previous studies\Cite{Kang2019,Benalcazar2017a,Benalcazar2017b}, EQMs of a finite-size system are obtained as the response of the (free) energy to quadrupolar electric field.
The difficulty lies in the fact that they generally depend on surface details\Cite{Trifunovic2019}, and thus a boundary contribution must be somehow subtracted.
To extract bulk properties, we consider the free energy of a small \textit{subsystem} deep inside the bulk and evaluate its response to the quadrupolar electrcic field. This idea is based on local thermodynamics.

Local thermodynamics has been utilized to define magnetic multipole moments.
After the pioneering work defining the orbital magnetization\Cite{Shi2007}, it has been recently adopted to define MQMs\Cite{Gao2018a,Shitade2018a,Gao2018b,Shitade2019}.
In earlier formulations, MQMs turned out to be unit-cell dependent\Cite{Ederer2007,Spaldin2008} or gauge-dependent\Cite{Batista2008,Thole2016,Chen2018}, which is undesirable for characterizing materials.
This problem can be circumvented with a thermodynamic approach, where MQMs are defined as coefficients of the free energy density in terms of the spatial gradient of magnetic field.
In addition, using the thermodynamic relation, a direct relationship of MQMs with the magnetoelectric coefficients has been proved\Cite{Gao2018a,Shitade2018a}.
The obtained formulas have already been used to estimate MQMs of materials\Cite{Gobel2019}.
Considering the success achieved by the thermodynamic approach, it is natural to ask whether it can define EQMs as well.

In this paper, we discuss a thermodynamic definition of EQMs to characterize the $C_4$ symmetry breaking of materials.
Following the orbital magnetization and MQMs, EQMs are defined by coefficients of the free energy density in terms of the spatial gradient of electric field.
In a way similar to MQMs\Cite{Gao2018a,Shitade2018a,Gao2018b,Shitade2019}, chemical potential derivative of the thermodynamic EQMs coincides with the electric susceptibility in insulators at zero temperature.

Thermodynamic EQMs have desirable properties as the order parameter of the $C_4$-symmetry breaking.
First, the obtained formulas are not only gauge invariant but also are unit-cell independent.
The latter property makes clear contrast to a simple option of taking average of the atomic-scale EQMs over a specific choice of the unit cell.
Next, thermodynamic EQMs have the advantage of wide applicability.
They are applicable to both insulators and metals at arbitrary temperature.
Furthermore, the framework is in principle not limited to electron systems, as long as local thermodynamics and the coupling to electric field can be introduced.
These features of thermodynamic EQMs are advantageous for a systematic characterization of $C_4$-breaking properties of materials via first-principles calculations.

The remaining part of the paper is constructed as follows.
In \Sec{sec:Derivation_of_EQM}, we discuss the setup and derive the formulas for EQMs.
Electric field is introduced as a scalar potential existing in a sufficiently small region compared with the system size.
\Sec{sec:model_calculation} is devoted to illustrate the obtained formulas for EQMs with a model of iron-based superconductors La$_{1-x}$F$_x$FeAsO.
An expression for general two-band systems is also shown.
In \Sec{sec:Summary}, we summarize the obtained results and make some comments.

\section{Derivation of EQM}
\label{sec:Derivation_of_EQM}
\subsection{Setup}
Let us consider a $d$-dimensional system under inhomogeneous electric field.
The Hamiltonian of the system is given by
\begin{equation}
\hat{H}=\hat{H}_0+\int d^dx\,(\phi(\bm{x})-\xi)\hat{n}(\bm{x}).\label{eq:Hamil}
\end{equation}
Here, $\xi$ is the electrochemical potential, while $\phi(\bm{x})$ is a scalar potential with the electron charge set to unity.
The unperturbed Hamiltonian $\hat{H}_0$ is assumed to be noninteracting in this paper, although a short-ranged interaction can be included to the formulation of local thermodynamics\Cite{Cooper1997}.
Thus, in the following, \Eq{eq:Hamil} represents the single-particle Hamiltonian, where the one-body particle-density operator $\hat{n}(\bm{x})=\sum_{s=\uparrow,\downarrow}\ket{\bm{x},s}\bra{\bm{x},s}$
is written with the spin-$s$ position eigenstates $\ket{\bm{x},s}$.

We assume that spatial modulation of $\phi(\bm{x})$ exists only in a small region compared with the system size $L^d$.
This allows us to consider thermal equilibrium of the system, whose density matrix is given by the Gibbs ensemble for the Hamiltonian~\eqref{eq:Hamil}.
In this situation, the scalar potential $\phi(\bm{x})$ would correspond to an inhomogeneous distribution of disorders, structural asymmetry, contact with a substrate, and so on, rather than an applied electric field.
Although electrons experience the force by the local electric field $-\nabla \phi(\bm{x})$, it is balanced with the statistical force $-\nabla\mu(\bm{x})$, with $\mu(\bm{x})$ being the chemical potential, as seen below.
Thereby, a static charge distribution is realized.
We also assume that the length scale $q^{-1}$ of $\phi(\bm{x})$ is larger than the decay length $l$ of Green's function.
This means that the whole system can be divided into weakly interacting small subsystems as a first approximation, allowing us to consider local thermodynamics.

Following the definitions of magnetic multipole moments~\cite{Shi2007,Gao2018a,Shitade2018a,Gao2018b,Shitade2019}, we define thermodynamic electric multipole moments by the variation of the free energy density:
\begin{widetext}
\begin{equation}
  d{F}(\bm{x})=\rho(\bm{x})d\phi(\bm{x})+p^i(\bm{x})d[\partial_i\phi(\bm{x})]+Q^{ij}(\bm{x})d[\partial_i\partial_j\phi(\bm{x})]+O(d[\nabla^3\phi(\bm{x})],[d\phi(\bm{x})]^2).\label{eq:thermo}
\end{equation}
\end{widetext}
Thermodynamic electric monopole, dipole, and quadrupole moments are defined by $\rho(\bm{x})$, $p^i(\bm{{x}})$, and $Q^{ij}(\bm{x})$, respectively.

In order to calculate the free energy density, we follow Ref.~\cite{Shi2007}: we first calculate energy density $E(\bm{x})$,
\begin{equation}
E(\bm{x})=\braket{\hat{\mathcal{H}}(\bm{x})},\quad \hat{\mathcal{H}}(\bm{x})\equiv\frac{1}{2}\{\hat{H},\hat{n}(\bm{x})\},\label{eq:def_E}
\end{equation}
and then obtain $F(\bm{x})$ from the thermodynamic relation
\begin{subequations}
\begin{gather}
\partial_\beta(\beta F(\bm{x}))=E(\bm{x}),\\
\lim_{\beta\to\infty} [F(\bm{x})-E(\bm{x})]=0,
\end{gather}
\end{subequations}
with $\beta=1/T$ being the inverse temperature.
To be precise, coarse graining is necessary, since the expression of $E(\bm{x})$ in \Eq{eq:def_E} involves contributions oscillating with a lattice period.
We explicitly show the coarse graining procedure in Appendix~\ref{app:app1}.

Since $\phi(\bm{x})$ varies slowly in space, the
system around position $\bm{x}$ is well approximated with the uniform Hamiltonian with the chemical potential $\xi-\phi(\bm{x})$.
By taking this state as the starting point, we calculated corrections to $E(\bm{x})$ in terms of spatial gradient of $\phi(\bm{x})$, following the spirit of gradient expansion\Cite{Rammer2007}.
The small parameter is given by $ql$, with the decay length $l$ originating from either excitation gap in insulators or thermal fluctuations for metals in the current setup (in reality, $l$ of the metals will be determined by impurity scattering at low temperature).
The result is given by [see Appendix~\ref{app:app1} for details],
\begin{align}
F(\bm{x})&=F_0(\xi-\phi(\bm{x}))+Q^{ij}(\xi-\phi(\bm{x}))\partial_i\partial_j\phi(\bm{x})\notag\\
&\ \ \,-\frac{1}{2}\frac{\partial Q^{ij}(\xi-\phi(\bm{x}))}{\partial\xi}\partial_i\phi(\bm{x})\partial_j\phi(\bm{x})+O(ql)^3.
\label{eq:eq4}
\end{align}
When the first term is retained, and the other terms of $O(ql)^2$ are neglected, $F(\bm{x})$ coincides with the free energy density of $\hat{H}_0$ with the chemical potential $\mu(\bm{x})=\xi-\phi(\bm{x})$, satisfying the condition for equilibrium under the inhomogeneous scalar potential\Cite{Cooper1997}.

For our purpose, it is sufficient to take the linear order in $\phi(\bm{x})$.
Rewriting $\phi(\bm{x})$ as $d\phi(\bm{x})$, we obtain
\begin{equation}
  dF(\bm{x})=\rho_0(\xi)d\phi(\bm{x})+Q^{ij}(\xi)d[\partial_i\partial_j\phi(\bm{x})],\label{eq:free_energy_result}
\end{equation}
up to $O(d\phi(\bm{x})^2,(ql)^3)$.
This result is reproduced with the Kubo formula, by calculating the responce of the energy density to the scalar potential (see Appendix~\ref{app:kubo}).
The first term trivially gives the electric charge $\rho_0(\xi)=-\partial F_0(\xi)/\partial\xi$.
The second term gives EQMs, the main topic of this paper. The details are discussed below.
Note that the electric dipole moment vanishes in the thermodynamic definition, because the thermodynamic approach is gauge-invariant. It is known that the electric dipole moment given by the Berry phase formula~\cite{Vanderbilt1993,KingSmith1993,Vanderbilt2018} is gauge-dependent
(gauge-invariant only modulo a lattice period, to be precise).
More in general, odd-rank electric multipole moments vanish within the thermodynamic approach (see Appendix~\ref{app:kubo}).
Coherent treatment of both odd- and even-rank electric multipole moments is left as a future issue.

\subsection{Formulas for EQMs}
In this subsection and \Sec{sec:model_calculation}, we focus on the properties of the system without a scalar potential $\phi(\bm{x})$.
In this case, the electrochemical potential $\xi$ is identified with the chemical potential $\mu$. Therefore, we use $\mu$  instead of $\xi$.

The expression for thermodynamic EQMs $Q^{ij}(\mu)$ is obtained as
\begin{widetext}
\begin{equation}
Q^{ij}(\mu)=\sum_{n}\int_{\mathrm{BZ}}\frac{d^dk}{(2\pi)^d}\left[\frac{1}{2}g_n^{ij}(\bm{k})f(\epsilon_n(\bm{k}))-X_n^{ij}(\bm{k})\int_{\epsilon_n(\bm{k})}^\infty d\epsilon\,f(\epsilon)-\frac{1}{12}m^{-1}_n(\bm{k})^{ij}f'(\epsilon_n(\bm{k}))\right].\label{eq:Result2}
\end{equation}
\end{widetext}
Here, $g_n^{ij}(\bm{k})$ is the quantum metric\Cite{Provost1980,Resta2011}
\begin{equation}
  g_n^{ij}(\bm{k})=\sum_{m\neq n}A_{nm}^i(\bm{k})A_{mn}^j(\bm{k})+\mathrm{c.c.},
  \label{eq:quantum_metric}
\end{equation}
while $X_n^{ij}(\bm{k})$ is given by
\begin{equation}
  X_n^{ij}(\bm{k})=-\sum_{m\neq n}\frac{A_{nm}^i(\bm{k})A_{mn}^j(\bm{k})+\mathrm{c.c.}}{\epsilon_n(\bm{k})-\epsilon_m(\bm{k})},
  \label{eq:Xnij}
\end{equation}
and $m^{-1}_n(\bm{k})^{ij}=\partial_{k_i}\partial_{k_j}\epsilon_n(\bm{k})$ is the inverse effective mass tensor.
For simplicity, we assumed that all bands are isolated from each other.
The expression which is valid in the presence of band touchings is shown in Appendix~\ref{app:app3}.
We adopted the following notations
\begin{subequations}
\begin{gather}
(\hat{H}_0-\mu)\ket{\psi_n(\bm{k})}=\epsilon_n(\bm{k})\ket{\psi_n(\bm{k})},\\
\ket{u_n(\bm{k})}=e^{-i\bm{k}\cdot\hat{\bm{x}}}\ket{\psi_n(\bm{k})},\\
A_{nm}^i(\bm{k})=-i\braket{u_n(\bm{k})|\partial_{k_i}u_m(\bm{k})},
\end{gather}
\end{subequations}
where wave number $\bm{k}$ takes a value in the first Brillouin zone (BZ).
It should be noted that $Q^{ij}(\mu)$ is gauge-invariant and also independent of unit-cell choices, since it is written with $\ket{u_n(\bm{k})}$.
In particular, the latter property is a direct consequence of the coarse graining.

Let us discuss the physical meaning of each term in thermodynamically defined EQMs, \Eq{eq:Result2}.
The meaning of the first term becomes most evident in insulators at zero temperature:
\begin{equation}
\int\frac{d^dk}{(2\pi)^d}g_n^{ij}(\bm{k})=\braket{w_n(0)|\hat{x}_i(1-\hat{P}_n)\hat{x}_j|w_n(0)},\label{eq:Wannier}
\end{equation}
where $\ket{w_n(0)}$ is the Wannier function of the $n$-th band at the home unit cell, while $\hat{P}_n$ is the projection operator onto the $n$-th band.
Equation~\eqref{eq:Wannier} is nothing but the gauge-invariant part of the EQM of Wannier functions\Cite{Marzari2012}.
Recently, it has been pointed out that $g_n^{ij}(\bm{k})$ can be interpreted as the EQM of electron wave packets\Cite{Lapa2019,Gao2019}.
Thus, \Eq{eq:Wannier} can also be viewed as the net EQM of all the occupied electron wave packets.

The second term of \Eq{eq:Result2} comes from the positional shift of electrons within each unit cell induced by an electric field\Cite{Gao2014}.
For insulators at zero temperature, this term contributes to the (free) energy density $\sim Q_{ij}\partial_i\partial_j\phi(\bm{x})$ by
\begin{equation}
  -\partial_i\left[X_n^{ij}(\bm{k})(-\partial_j\phi(\bm{x}))\right]\cdot\epsilon_n(\bm{k}).\label{eq:pol_charge}
\end{equation}
Note that \Eq{eq:pol_charge} has the form of the polarization charge multiplied with $\epsilon_n(\bm{k})$.
Actually, the correction of the Berry connection ($\sim$ polarization) by $\phi(\bm{x})$ is given by $X_n^{ij}(\bm{k})(-\partial_j\phi(\bm{x}))$\Cite{Gao2014}.
Thus, the second term of \Eq{eq:Result2} reflects local increase of electrons
with energy $\epsilon_n(\bm{k})$.
This point is closely related to the thermodynamic relation, which is discussed soon later.

The appearane of the effective mass tensor in the third term of \Eq{eq:Result2} can be understood from the density oscillations of noninteracting electron gas, simply caused by the mixing of plane waves with slightly different wave numbers.
Actually, the term $-f''(\epsilon)/12 m$ also appears in the expansion of the electric susceptibility of electron gas, i.e. the Lindhard function (see Appendix~\ref{app:lindhard}).

\subsection{Thermodynamic relations and polarization charge}
As a virtue of thermodynamic formulation, there exists a thermodynamic relation between EQMs and the electric susceptibility.
Here, we show the derivation from the perspective of thermodynamics, following Refs.~\cite{Gao2018a,Shitade2019,Gao2018b,Shitade2018a}.
The result can also be obtained by directly calculating the charge density $\rho(\bm{x})$ from \Eq{eq:eq4} (see Appendix~\ref{app:ThermoRel}).

Let us consider insulators at zero temperature.
According to the expression of the free energy density~\eqref{eq:thermo}, the Maxwell relation holds between EQMs and the charge density:
\begin{equation}
\frac{\partial Q^{ij}(\bm{x})}{\partial\phi(\bm{x})}=\frac{\partial \rho(\bm{x})}{\partial (\partial_i\partial_j\phi(\bm{x}))}.\label{eq:thermodynamic_relation}
\end{equation}
The left-hand side is equated with $-\partial Q^{ij}(\mu-\phi(\bm{x}))/\partial\mu\to-\partial Q^{ij}(\mu)/\partial\mu$ for $\phi(\bm{x})\to0$.
As for right-hand side, we use the fact that the electron density is given only by the polarization charge
\begin{equation}
  \rho(\bm{x})-\rho_0=-\partial_i(\chi_{ij}^eE_j(\bm{x}))=\chi_{ij}^e\partial_i\partial_j\phi(\bm{x}),
\end{equation}
for insulators at zero temperature.
Here, $\chi_{ij}^e$ is the electric susceptibility, while $E_j(\bm{x})=-\partial_j\phi(\bm{x})$ is an electric field.
Thus, we obtain the thermodynamic relation
\begin{equation}
\frac{\partial Q^{ij}(\mu)}{\partial\mu}=-\chi_{ij}^e.\label{eq:thermo_rel}
\end{equation}
This relation should hold even in the presence of a short-range interaction beyond non-interacting cases.

The thermodynamic relation~\eqref{eq:thermo_rel} holds for the obtained EQMs.
Actually, the electric susceptibility is given by the well-known expression (see e.g. \Ref{Claudio1995})
\begin{equation}
  \chi_{ij}^e=\sum_{\epsilon_n<0}\int\frac{d^dk}{(2\pi)^d}\,X_n^{ij}(\bm{k}),
  \label{eq:electric_susceptibility}
\end{equation}
for noninteracting insulators.
This is equal to $-\partial Q^{ij}(\mu)/\partial\mu$ calculated from \Eq{eq:Result2}.
Equation~\eqref{eq:thermo_rel} is the analogue of the thermodynamic relations between the orbital magnetization and the Hall conductivity\Cite{Streda1982,Ceresoli2006,Xiao2010}, as well as between MQMs and the magnetoelectric polarizability\Cite{Gao2018a,Shitade2019,Gao2018b,Shitade2018a}.

\section{Model calculation}
\label{sec:model_calculation}
Here, we illustrate the formula \Eq{eq:Result2} for EQMs with a simple two-band model of an iron-based superconductor.
Before going to the detail, we show general expressions for two-band systems to get an intuition.
Let us consider a two-band Bloch Hamiltonian $H(\bm{k})=h_0(\bm{k})+\bm{h}(\bm{k})\cdot\bm{\sigma}$.
In this case, coefficients of \Eq{eq:Result2} are simplified for the lower band as follows:
\begin{gather}
  g_-^{ij}(\bm{k})=\frac{1}{2}\partial_{k_i}\hat{h}(\bm{k})\cdot\partial_{k_j}\hat{h}(\bm{k}),\\
X_-^{ij}(\bm{k})=\frac{1}{4h(\bm{k})}\partial_{k_i}\hat{h}(\bm{k})\cdot\partial_{k_j}\hat{h}(\bm{k}),
\end{gather}
with $\hat{h}(\bm{k})=\bm{h}(\bm{k})/h(\bm{k})$ and $h(\bm{k})=|\bm{h}(\bm{k})|$\Cite{Gao2014}.
Thus, for the insulator at zero temperature, the EQM is simply given by
\begin{equation}
  Q_{ij}(\mu)=\frac{1}{4}\int\frac{d^dk}{(2\pi)^d}\,\frac{h_0(\bm{k})}{h(\bm{k})}\partial_{k_i}\hat{h}(\bm{k})\cdot\partial_{k_j}\hat{h}(\bm{k}).\label{eq:2band}
\end{equation}
As is evident from the expression, a large contribution comes from the points where the band gap $2h(\bm{k})$ is small.
For metals, $Q_{ij}(\mu)$ is obtained by restricting the $\bm{k}$ integral to the Fermi sea, along with the inverse effective mass tensor averaged on the Fermi surface.
It is natural to expect a large contribution from the Fermi surfaces, while there is no reason to expect a small contribution from the Fermi sea terms.
Actually, we find that contributions from the Fermi surface and Fermi sea are comparable in the model shown below.

\begin{figure*}
  \begin{tabular}{ll}
    (a)&(b)\\
\includegraphics[width=0.55\textwidth]{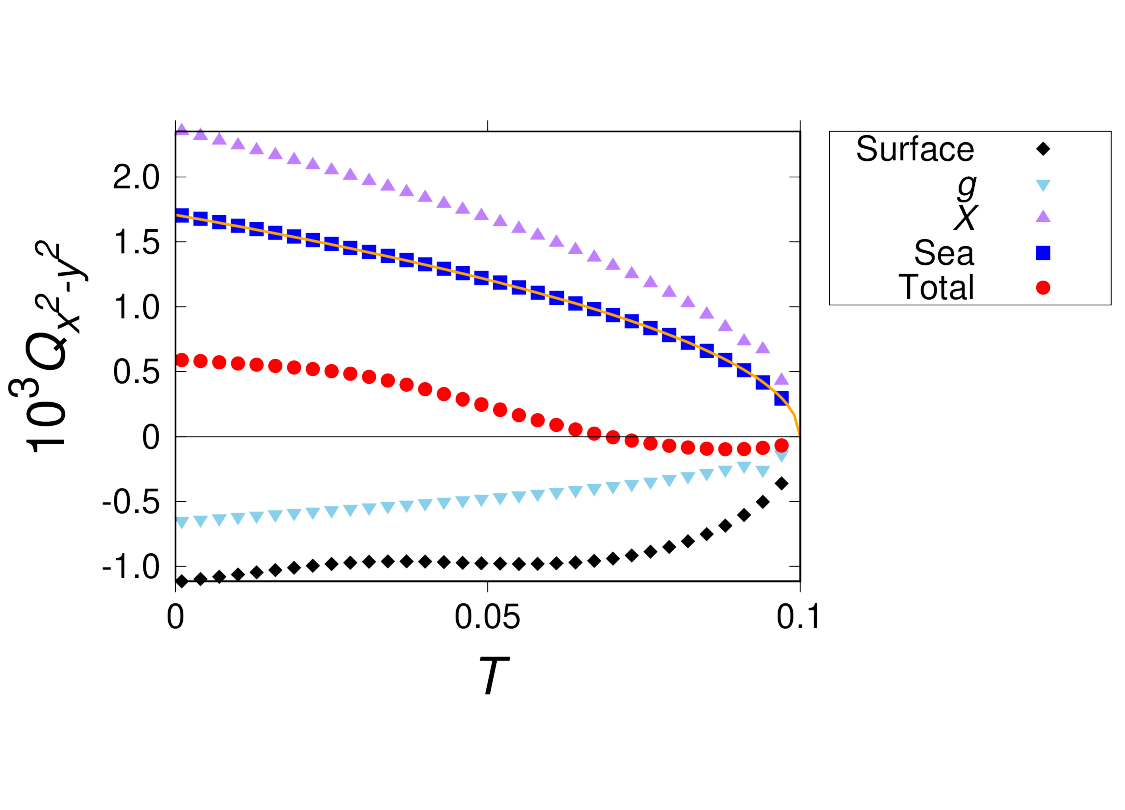} \hspace*{5mm}&
    \includegraphics[width=0.37\textwidth]{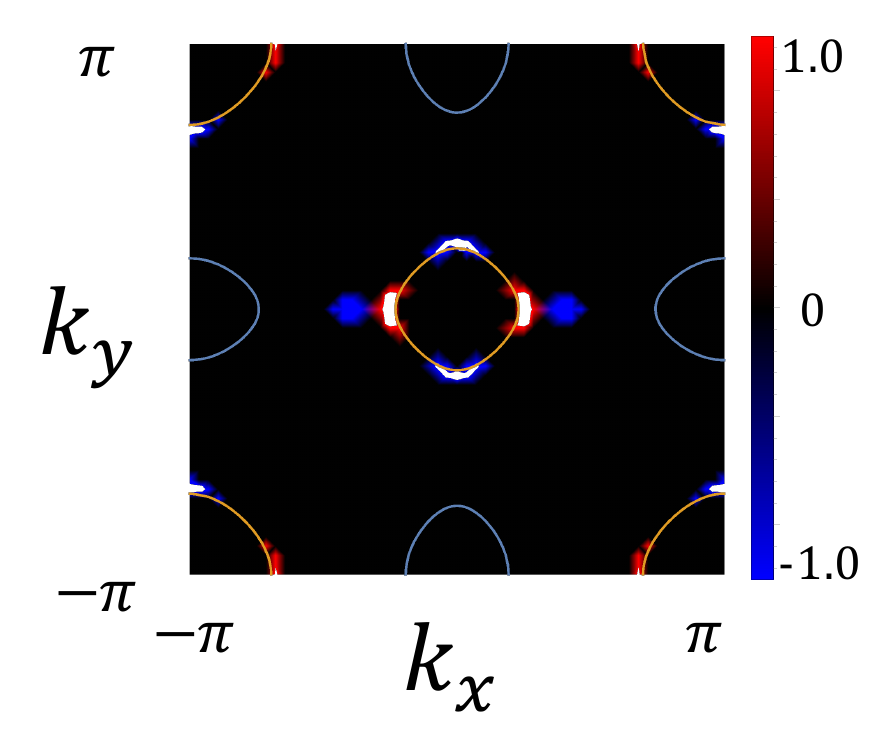}
\end{tabular}
  \caption{(a) Temperature dependence of EQM of the model~\eqref{eq:RSmodel} for LaFeAsO$_{1-x}$F$_x$.
  Total EQM is plotted with red dots, while the Fermi-surface and Fermi-sea contributions are plotted with black closed rhombuses and blue closed squares, respectively.
  We also show the contributions to the Fermi-sea terms from the quantum metric and $X_n^{ij}(\bm{k})$ terms by skyblue lower and purple upper triangles, respectively.
  The orange line (overlapping with blue squares) shows a fitting curve, $0.054\Delta(T)$.
  The $\bm{k}$ mesh was taken for $dk=2\pi/L_1$ for $\bm{k}$ points away from the Fermi surfaces, with $L_1=\mathrm{max}(300,(150\beta)^{2/3})$.
  $\bm{k}$ points near the Fermi surfaces are evaluated with $\bm{k}$ mesh $dk'=\mathrm{min}(dk,2\pi/150\beta)$.
(b) Contributions to the interband component from each $\bm{k}$ point for $T=0.05$.
Orange lines show the hole Fermi surfaces, while blue lines show the electron ones, for $\Delta=0$.
  White region corresponds to $\bm{k}$ points contributing to the EQM with absolute values larger than $1.0$.
  }
  \label{fig:EQM}
\end{figure*}

Let us focus on an iron-based superconductor LaFeAsO$_{1-x}$F$_x$\Cite{Kamihara2008,Stewart2011,Dai2015}.
This compound is known as the first discovered high-$T_{\mathrm{c}}$ iron-based superconductor, showing superconductivity for $x\ge 0.04$ with the maximum transition temperature $T_{\mathrm{c}}=26$~K\Cite{Kamihara2008}.
In the parent compound LaFeAsO, there is a structural phase transition at $T_{\mathrm{s}}\sim 160~\mathrm{K}$, followed by a spin density wave (SDW) transition at $T_{\mathrm{N}}\sim 140~\mathrm{K}$\Cite{Nomura2008,Cruz2008}.
The symmetry of the crystal lowers from tetragonal (space group $P4/nmm$) to orthorhombic (space group $Cmma$) with breaking the $C_4$ symmetry\Cite{Nomura2008,Stewart2011}.
In the following, we illustrate the formula for EQMs with a minimal model of LaFeAsO.
The $C_4$-breaking structural phase transition is treated phenomenologically, while the effect of SDW is neglected for simplicity.

A minimal description of LaFeAsO is achieved by the two-band model proposed by Raghu \textit{et al}.\Cite{Raghu2008},
\begin{subequations}
\begin{gather}
  h_0(\bm{k})=\epsilon_+(\bm{k})-\mu,\quad h_x(\bm{k})=\epsilon_{xy}(\bm{k}),\\
  h_y(\bm{k})=0,\quad h_z(\bm{k})=\epsilon_-(\bm{k})+\Delta,
\end{gather}
with
\begin{align}
  \epsilon_+(\bm{k})&=-(t_1+t_2)(\cos k_x+\cos k_y)\notag\\
  &\qquad\qquad-4t_3\cos k_x\cos k_y,\\
  \epsilon_-(\bm{k})&=-(t_1-t_2)(\cos k_x-\cos k_y),\\
  \epsilon_{xy}(\bm{k})&=-4t_4\sin k_x\sin k_y.
\end{align}
\label{eq:RSmodel}
\end{subequations}
Here, we additionally introduced $\Delta$ as the simplest order parameter for the $C_4$-symmetry breaking.
Physically, $\Delta$ describes an orbital order, since the index $\sigma_z=\pm1$ represents $d_{xz}$ and $d_{yz}$ orbital states.
The parameters of the model are given by\Cite{Raghu2008}
\begin{equation}
    (t_1,\,t_2,\,t_3,\,t_4,\,\mu)=(-1.0,1.3,-0.85,-0.85,1.45),
\end{equation}
qualitatively reproducing the Fermi surfaces of La$_{1-x}$F$_x$FeAsO obtained by first-principle calculations\Cite{Singh2008,Xu2008,Mazin2008,Haule2008}.
There are hole Fermi surfaces around $(k_x,k_y)=(0,0)$ and $(\pi,\pi)$, in addition to the electron Fermi surfaces around $(0,\pi)$ and $(\pi,0)$ (\Fig{fig:EQM}~(b)).
In the following, we evaluate $Q_{x^2-y^2}=Q_{xx}-Q_{yy}$, since $Q_{xy}=0$ due to the mirror symmetries.

In order to calculate the temperature dependence of the EQM, we assume for simplicity
\begin{equation}
\Delta(T)=\begin{cases}0&(T>T_{\mathrm{s}})\\
\Delta_0\sqrt{1-T/T_c}&(T\le T_{\mathrm{s}}),
\end{cases}
\end{equation}
with $\Delta_0=T_{\mathrm{s}}=0.1\,|t_1|$.
The result is shown in \Fig{fig:EQM}~(a).
$Q_{x^2-y^2}$ is shown by red dots, while contributions from the first, second, and third terms in \Eq{eq:Result2} are shown by skyblue, purple, and black points, respectively.
We call the sum of the former two as the Fermi-sea contribution $Q_{x^2-y^2}^{\mathrm{sea}}$, and show it by blue rectangles.
The contribution from the third term (inverse effective mass tensor) is called Fermi-surface contribution, in the following.

The Fermi-sea component scales linearly with $\Delta(T)$, fitted well with $Q_{x^2-y^2}^{\mathrm{sea}}\simeq 0.054\Delta(T)$ (orange line in \Fig{fig:EQM}~(a)).
This is because the band gap is sufficiently large at every $\bm{k}$ points contributing to the integral (see also the discussion below).
Note that a hump structure appears around $T=0.09$ in the contributions from $g_n^{ij}(\bm{k})$ and $X_n^{ij}(\bm{k})$. %
This is probably related to the band degenerate points, around which $g_n^{ij}(\bm{k})$ and $X_n^{ij}(\bm{k})$ tend to diverge with the opposite sign.
However, their summation is always regular, and therefore no anomaly is observed in the total Fermi-sea contribution.
We also find a shoulder structure in the Fermi-surface component, leading to the sign reversal of the total EQM $Q_{x^2-y^2}$. This structure originates from the competition of the following two effects: On one hand, the inverse effective mass tensor $1/m_{xx}-1/m_{yy}$ has peaks slightly off the Fermi level as a function of energy. As temperature increases, they can contribute to the Fermi-surface component owing to the thermal broadening of $f^{\prime}(\epsilon_n(\bm{k}))$. On the other hand, they shrink towards $T_{\rm c}$ together with the order parameter $\Delta(T)$. Overall, both the Fermi-surface and Fermi-sea components make comparable contributions to $Q_{x^2-y^2}$ with different signs.

For the purpose of considering the origin of the large Fermi-sea contribution, we also show the $\bm{k}$ dependence of the first two terms of \Eq{eq:Result2} in \Fig{fig:EQM}~(b).
Interestingly, $Q_{x^2-y^2}^{\mathrm{sea}}$ is largely contributed from the vicinity of the Fermi surfaces surrounding $\bm{k} = (0,0)$ and $(\pi,\pi)$, not from that of these points themselves. The reason is as follows: At these points, $h(\bm{k})$ is small, and $X_n^{ij}(\bm{k})$ becomes large. However, considering that $X_n^{ij}(\bm{k})$ is antisymmetric between the two bands, the distribution factor is approximated as $f(h_0 + h) - f(h_0 - h)$ after the summation over $n$ and becomes vanishingly small. As a result of this competition,
the integrand is largest near the Fermi surfaces.
In this way, it is generally expected that there appears a large contribution from the Fermi-sea term when small Fermi surfaces enclose (nearly) degenerate points.

\begin{figure}
    \centering
      \includegraphics[width=0.5\textwidth]{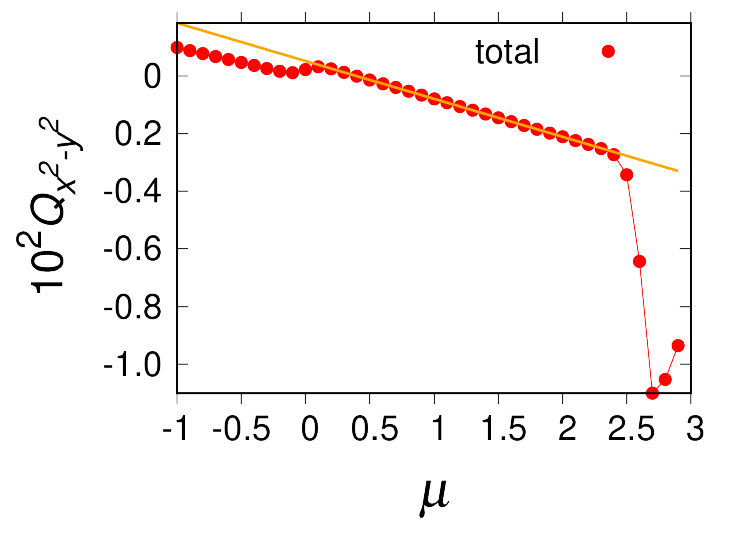}
  \caption{Chemical potential dependence of the EQM of the model \eqref{eq:model2} at $T=0.05$.
  The $\bm{k}$-mesh is the same as \Fig{fig:EQM}~(a).
  The red dots show the total EQM, while the orange line shows a fitting line, $-1.31874\times 10^{-3}(\mu-1)-7.94151\times 10^{-4}$.
  The red line is the guide for the eye.}
  \label{fig:EQM2}
\end{figure}

Next, we verify the thermodynamic relation for insulators at zero temperature.
For the purpose of describing an insulating state, we add a spin-orbit coupling $\lambda=4.0$ to the previous model by
\begin{align}
  H'(\bm{k})=(h_0(\bm{k})+\bm{h}(\bm{k})\cdot\bm{\sigma})\otimes s_0+\lambda\sigma_y\otimes s_z.
  \label{eq:model2}
\end{align}
The unrealistically large value of $\lambda$ is just a technical mean to obtain a model of an insulator.
Here, $s_\mu$ is the Pauli matrices for the spin degree of freedom.
We show the chemical potential dependence of the EQM in \Fig{fig:EQM2}, where the system is insulating around $\mu=1$.
For a technical reason we choose $T=0.05$, which is sufficiently smaller than the insulating gap.
A linear change of the EQM in the insulating region is illustrated, whose slope is well approximated by $-\chi_{ij}^e$ with the calculated electric susceptibility $\chi_{ij}^e=1.31874\times 10^{-3}$.
Thus, we have confirmed that the thermodynamic relation \Eq{eq:thermodynamic_relation} holds.

\section{Summary and Discussion}
\label{sec:Summary}
In this paper, we proposed a thermodynamic definition of EQMs. %
The obtained formulas are gauge invariant, unit-cell independent, and applicable to both insulators and metals at zero and nonzero temperature.
A thermodynamic relation connecting EQMs with electric susceptibility is obtained.
The results are confirmed with a numerical calculation of a model for an iron-based superconductor LaFeAsO$_{1-x}$F$_{x}$.
We argue that the thermodynamic EQMs can serve as a common measure of the $C_4$-symmetry breaking in quantum phases.

As a concluding remark, we discuss the relation of the thermodynamic EQMs with previous studies.
We first note that topological corner charge\Cite{Benalcazar2017a,Benalcazar2017b} is not captured by thermodynamic EQMs, which are gauge invariant and thus always vanish in the presence of the $C_4$ symmetry.
Our interest is more on the nontopological contribution to the quadrupolar properties of materials.
We believe that an important contribution is captured with thermodynamic EQMs.

We also refer to several recent works studying EQM-related properties of materials.
Attempts have been made to evaluate the corner charge via Wannier functions without imposing rotational symmetries\Cite{Zhou2015,Trifunovic2020}.
The corner charge is divided into the contributions from the quadrupole tensor (the expectation value of e.g. $\hat{x}\hat{y}$ by Wannier functions) and the edge polarization, by selecting the bulk Wannier functions from the localized basis set in systems with open boundary conditions.
As noted in this paper, it is known that the gauge-invariant part of the quadrupolar tensor is equivalent with the quantum metric\Cite{Marzari2012},
which is also included in the thermodynamic EQMs.
Recently it is shown via wave-packet theory that the quantum metric naturally appears in the gradient expansion of the charge density in insulators at zero temperature\Cite{Zhao2020}.
We expect that there is a complementary relation between the quantum metric and the thermodynamic EQMs to describe polarization-related properties of materials, as is the case for the current and magnetization responses (related to orbital magnetization\Cite{Xiao2010} and thermodynamic MQMs\Cite{Gao2018a,Shitade2018a,Gao2018b,Shitade2019}).
Comprehensive understanding of these EQM-related properties of materials is left as an important future issue.

\begin{acknowledgments}
  \textit{Acknowledgments ---}
We thank J. Ishizuka and T. Kitamura for fruitful discussions.
This work was supported by KAKENHI (Grants No. JP15H05884, No.~JP17J10588, No. JP18H04225, No. JP18H05227, No. JP18H01178, No. JP18K13508, and No. JP20H05159) from the Japan Society for the Promotion of Science (JSPS).
\end{acknowledgments}

%

\begin{widetext}
\appendix
\section{Derivation of thermodynamic multipole moments}
\label{app:app1}
\subsection{Expansion around a point of the system}
Here, we show details of the calculation of $F(\bm{x})$.
In the following, we drop hat notations from operators like $\hat{H}\to H$, for simplicity.
The energy density is given by
\begin{subequations}
\begin{align}
  \braket{\mathcal{H}(\bm{x})}&=\sum_\nu\braket{\nu|{\mathcal{H}}(\bm{x})|\nu}f(\epsilon_\nu)\\
&=\oint\frac{dz}{2\pi i}\Tr[{n}(\bm{x})G(z)H]f(z),\label{eq:Ex}
\end{align}
\end{subequations}
where $\ket{\nu}$ is the energy eigenstate satisfying $H\ket{\nu}=\epsilon_\nu\ket{\nu}$, $f(z)=(e^{\beta z}+1)^{-1}$ is the Fermi distribution function, and $G(z)=(z-H)^{-1}$ is the Green's function.
The integration contour runs above and below the real axis, as shown in Fig.~\ref{fig:contour}.
\begin{figure}[h]
  \centering
  \includegraphics[width=0.5\textwidth]{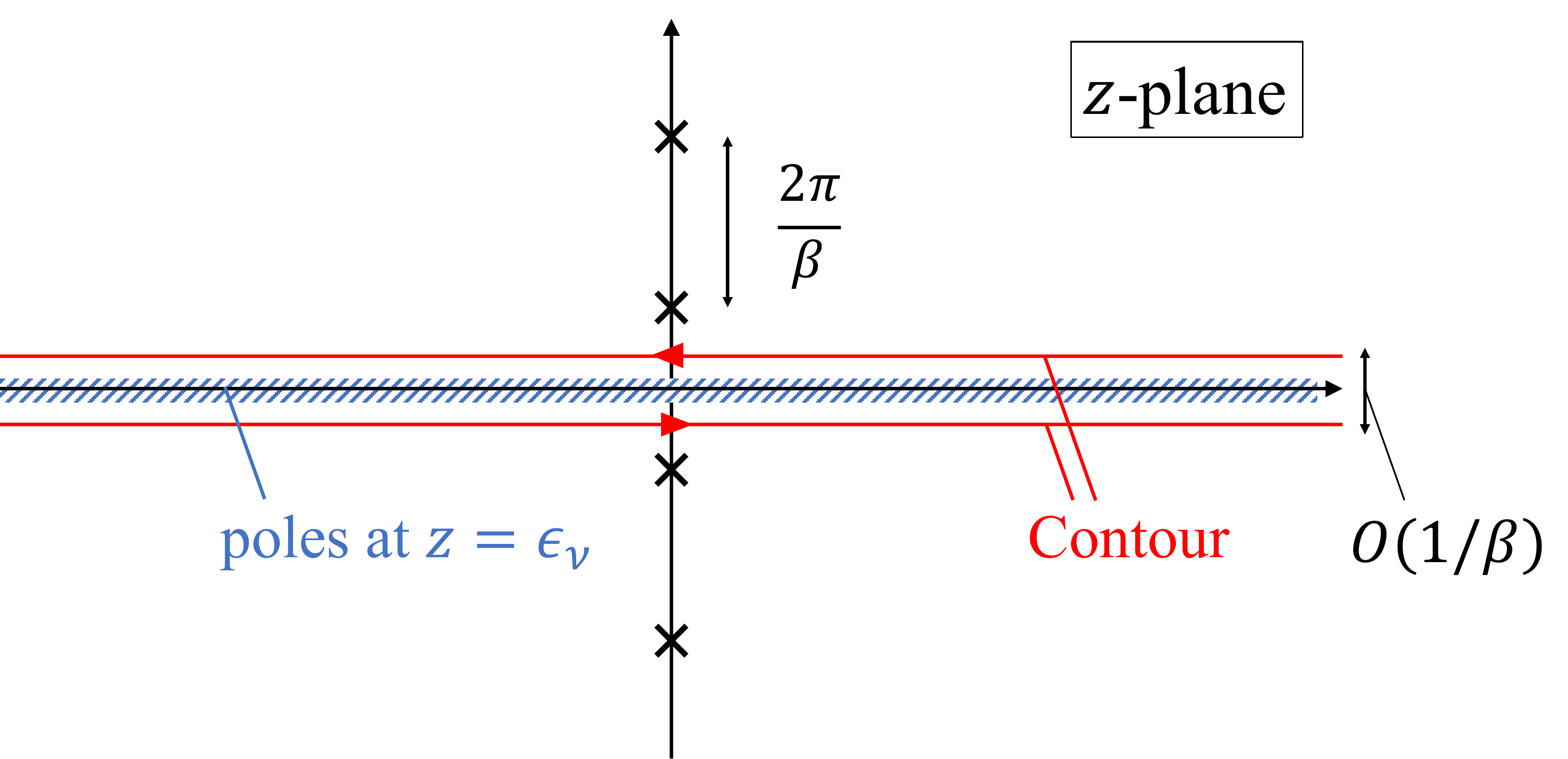}
  \caption{Contour of the complex integral \Eq{eq:Ex} and poles of the integrand.
  For insulators, poles are absent around $z=0$.%
  }
  \label{fig:contour}
\end{figure}

The idea of calculation is as follows.
Let us consider an arbitrary point around $\bm{x}$, and name it $\bm{x}'$.
Since $\phi(\bm{x})$ varies slowly in space, we can use $\phi(\bm{x})$ as the zero-th order approximation of $\phi(\bm{x}')$.
Therefore, it is natural to take \textit{Hamiltonian with constant scalar potential whose value is $\phi(\bm{x})$} as the unperturbed Hamiltonian, when we talk about physical properties of the system around $\bm{x}$.
Thus, we introduce `Hamiltonian around $\bm{x}$' by
\begin{equation}
  H(\bm{x})\equiv H_0(\xi)+\int d^dx'\,\phi(\bm{x})n(\bm{x}')=H_0(\xi-\phi(\bm{x})),
\end{equation}
instead of the original Hamiltonian
\begin{equation}
  H\equiv H_0(\xi)+\int d^dx'\,\phi(\bm{x}')n(\bm{x}').
\end{equation}
Let us write Green's function for $H_0(\xi)$ and $H(\bm{x})$ as $G_0(z)=(z-H_0(\xi))^{-1}$ and $G_{\bm{x}}(z)=(z-H(\bm{x}))^{-1}$, respectively, where $G_{\bm{x}}(z)=G_0(z)|_{\mu\to\mu-\phi(\bm{x})}$ holds.
The energy density is given by
\begin{subequations}
\begin{align}
  \braket{\mathcal{H}(\bm{x})}-\braket{\mathcal{H}_0(\bm{x})}_{\mu\to\mu-\phi(\bm{x})}
  &=\oint\frac{dz}{2\pi i}\Tr\Bigl[{n}(\bm{x})\bigl\{G(z)H-G_{\bm{x}}(z)H(\bm{x})\bigr\}\Bigr]f(z)\\
  &=\oint\frac{dz}{2\pi i}\,z\Tr\Bigl[{n}(\bm{x})\bigl\{G(z)-G_{\bm{x}}(z)\bigr\}\Bigr]f(z).
\end{align}
\end{subequations}
Now, the physical picture discussed above is correctly reproduced by the Dyson series:
\begin{subequations}
\begin{align}
  G(z)-G_{\bm{x}}(z)&=G_{\bm{x}}(z)(H-H(\bm{x}))G(z)\\
  &=G_{\bm{x}}(z)(H-H(\bm{x}))G_{\bm{x}}(z)+G_{\bm{x}}(z)(H-H(\bm{x}))G_{\bm{x}}(z)(H-H(\bm{x}))G_{\bm{x}}(z)+\cdots.\label{eq:expansion_G}
\end{align}
\end{subequations}
The right-handed side represents scattering processes due to finite gradient of the scalar potential:
\begin{subequations}
  \label{eq:expansion_dH}
\begin{align}
  H-H(\bm{x})&=\int d^dx'\,(\phi(\bm{x}')-\phi(\bm{x}))n(\bm{x}')\\
  &=\partial_i\phi(\bm{x})\int d^dx'\,(x'_i-x_i)n(\bm{x}')\\
  &\qquad+\frac{1}{2}\partial_i\partial_j\phi(\bm{x})\int d^dx'\,(x'_i-x_i)(x'_j-x_j)n(\bm{x}')+\cdots.
\end{align}
\end{subequations}
It should be emphasized that this expansion is regular.
Let us take one of the contributions as an example:
\begin{equation}
  (x'_i-x_i)\Tr[n(\bm{x})G_{\bm{x}}(z)n(\bm{x}')G_{\bm{x}}(z)]=\lim_{\mu\to\mu-\phi(\bm{x})}\sum_{s,s'}({x}'_i-{x}_i)G_{0ss'}(z,\bm{x},\bm{x}')G_{0s's}(z,\bm{x}',\bm{x}),
\end{equation}
with $G_{0ss'}(z,\bm{x},\bm{x}')=\braket{\bm{x},s|G_0(z)|\bm{x}',s'}$.
Note that $G_{0ss'}(z,\bm{x},\bm{x}')$ decays exponentially in space with some length scale $l$.
For insulators at low temperatures, $l$ is roughly estimated to be the localization length $\sim1/\sqrt{m^*\Delta}$ with $m^*$ the effective mass and $\Delta$ the insulating gap.
For (ideally clean, noninteracting) metals, $l$ is estimated to be the thermal length $\beta v_{\mathrm{F}}$, with $v_{\mathrm{F}}$ the Fermi velocity.
(These points become more evident by rewriting the expression in terms of Matsubara frequencies.)
By writing the length scale of $\phi(\bm{x})$ as $q^{-1}$, the expansions Eqs.~\eqref{eq:expansion_G} and \eqref{eq:expansion_dH} are in terms of $q(x_i-x'_i)\sim ql$.
Thus, our expansions are justified for $\phi(\bm{x})$ satisfying $ql\ll 1$.

In the following subsections, we explicitly evaluate the contribution to $\braket{\mathcal{H}(\bm{x})}$ up to $O(q^2)$, based on the expression:
\begin{subequations}
\label{eq:start}
\begin{align}
  &\braket{\mathcal{H}(\bm{x})}-\braket{\mathcal{H}_0(\bm{x})}_{\mu\to\mu-\phi(\bm{x})}\\
  &=\lim_{\mu\to\mu-\phi(\bm{x})}\oint\frac{dz}{2\pi i}\,zf(z)\Bigl(\partial_i\phi(\bm{x})p_i(z,\bm{x})+\partial_i\partial_j\phi(\bm{x})Q_{ij}(z,\bm{x})+\partial_i\phi(\bm{x})\partial_j\phi(\bm{x})Q_{ij}'(z,\bm{x})\Bigr)+O(q^3),
\end{align}
where
\begin{align}
p_i(z,\bm{x})&=-\int d^dx'\,(x_i-x_i')\Tr\Bigl[{n}(\bm{x})G_0(z)n(\bm{x}')G_{0}(z)\Bigr],\label{eq:p_zx}\\
  Q_{ij}(z,\bm{x})&=-\frac{1}{2}\int d^dx'\,(x_i-x'_i)(x_j'-x_j)\Tr\Bigl[{n}(\bm{x})G_0(z)n(\bm{x}')G_{0}(z)\Bigr],\label{eq:Q_zx}\\
  Q'_{ij}(z,\bm{x})&=-\int d^dx'\int d^dx''\,(x_i-x_i')(x_j''-x_j)\Tr\Bigl[{n}(\bm{x})G_0(z)n(\bm{x}')G_{0}(z)n(\bm{x}'')G_0(z)\Bigr]\label{eq:Qp_zx}.
\end{align}
\end{subequations}
As for $Q'_{ij}(z,\bm{x})$, only its symmetric part contributes to $\braket{\mathcal{H}(\bm{x})}$.
Therefore, in the follwoing, we rename $(Q'_{ij}(z,\bm{x})+Q'_{ji}(z,\bm{x}))/2$ of \Eq{eq:Qp_zx} as $Q'_{ij}(z,\bm{x})$ (used in \Eq{eq:Qp_temp}).

\subsection{Fourier transform}
Here, we show some details about Fourier transform of position eigenstates and Green's function, for the latter use.
Let us decompose $\bm{x}$ as follows:
\begin{equation}
\bm{x}=\bm{R}+\Delta\bm{r}.
\end{equation}
Here, $\bm{R}$ is the origin of the unit cell to which $\bm{x}$ belongs, and $\Delta\bm{r}$ is the position measured from the origin of the unit cell.
Then, Fourier transform of $\ket{\bm{x},s}$ is given by
\begin{subequations}
\begin{gather}
\ket{\bm{k},\Delta\bm{r},s}=\frac{1}{\sqrt{N}}\sum_{\bm{R}}e^{i\bm{k}\cdot(\bm{R}+\Delta\bm{r})}\ket{\bm{R}+\Delta\bm{r},s},\\
\ket{\bm{x},s}=\frac{1}{\sqrt{N}}\sum_{\bm{k}}e^{-i\bm{k}\cdot\bm{x}}\ket{\bm{k},\Delta\bm{r},s},
\end{gather}
\end{subequations}
where $N$ is the total number of unit cells, and $\bm{k}$ is in the first Brillouin zone.
Accordingly, Green's function is Fourier transformed by
\begin{subequations}
\begin{align}
[G_0(z,\bm{k})]_{(\Delta\bm{r},s),(\Delta\bm{r}',s')}&=\braket{\bm{k},\Delta\bm{r},s|G_0(z)|\bm{k},\Delta\bm{r}',s'}\\
&=\frac{1}{N}\sum_{\bm{R}}e^{-i\bm{k}\cdot(\bm{R}+\Delta\bm{r}-\Delta\bm{r}')}G_{0s,s'}(z,\bm{R}+\Delta\bm{r},\Delta\bm{r}'),\\
G_0(z,\bm{x},\bm{x}')&=\frac{1}{N}\sum_{\bm{k}}e^{i\bm{k}\cdot(\bm{x}-\bm{x}')}[G_0(z,\bm{k})]_{(\Delta\bm{r},s),(\Delta\bm{r}',s')}.
\end{align}
\end{subequations}
The following equality holds:
\begin{subequations}
\begin{align}
(x_i-x'_i)G_0(z,\bm{x},\bm{x}')&=\frac{1}{N}\sum_{\bm{k}}\left(-i\partial_{{k}_i}e^{i\bm{k}\cdot(\bm{x}-\bm{x}')}\right)[G_0(z,\bm{k})]_{(\Delta\bm{r},s),(\Delta\bm{r}',s')}\\
&=\frac{1}{N}\sum_{\bm{k}}e^{i\bm{k}\cdot(\bm{x}-\bm{x}')}i\partial_{{k}_i}[G_0(z,\bm{k})]_{(\Delta\bm{r},s),(\Delta\bm{r}',s')}
\end{align}
\end{subequations}
The first line is justified since $(x_i-x_i')\Delta{k}\sim 2\pi l/L\ll1$.
To obtain the second line, we dropped the total derivative because $e^{i\bm{k}\cdot(\bm{x}-\bm{x}')}[G_0(z,\bm{k})]_{(\Delta\bm{r},s),(\Delta\bm{r}',s')}$ has periodicity of the Brillouin zone, though $G_0(z,\bm{k})$ does not.

Since $G_0(z,\bm{k})$ has the matrix structure in the $(\Delta\bm{r},s)$ space, it is useful to define trace in this space.
We write this as ``$\tr$".
We also define projection operator in $(\Delta\bm{r},s)$ space onto the state with $\Delta\bm{r}$:
\begin{equation}
[P(\Delta\bm{r})]_{(\Delta\bm{r}_1,s_1),(\Delta\bm{r}_2,s_2)}=\delta(\Delta\bm{r}-\Delta\bm{r}_1)\delta(\Delta\bm{r}_1-\Delta\bm{r}_2)\delta_{ss'}.
\end{equation}
We also define the projection operator in the whole space with the same symbol:
\begin{equation}
P(\Delta\bm{r})=\sum_{\bm{R},s}\ket{\bm{R}+\Delta\bm{r},s}\bra{\bm{R}+\Delta\bm{r},s}=\sum_{\bm{R}}n(\bm{R}+\Delta\bm{r}).
\end{equation}
These operators appear in all the coefficients of \Eq{eq:start}.
Since $P(\Delta\bm{r})$ specifies microscopic position within each unit cell,
it should be traced out in the final expression of thermodynamic electric multipole moments.
This requires an explicit coarse graining, as we discuss in the next subsection.

\subsection{Coarse graining}
In order to illustrate the necessity of coarse graining, let us evaluate $\braket{\mathcal{H}_0(\bm{x})}$.
Since $H_0$ has a lattice periodicity, we obtain
\begin{subequations}
\begin{align}
\braket{\mathcal{H}_0(\bm{x})}&=\Tr[n(\bm{R}+\Delta\bm{r})Hf(H)]\\
&=\frac{1}{N}\sum_{\bm{R}'}\Tr[n(\bm{R}'+\Delta\bm{r})Hf(H)]\\
&=\frac{1}{N}\Tr[P(\Delta\bm{r})Hf(H)].
\end{align}
\end{subequations}
Thus, $\braket{\mathcal{H}_0(\bm{x})}$ does not coincide with the average energy density $\frac{1}{V}\Tr[Hf(H)]$.
Here, $V=N\Omega$ is the volume of the system and $\Omega$ is the volume of a unit cell.
In order to go along with the spirit of local thermodynamics, and to trace out microscopic oscillations, we introduce coarse graining as follows\Cite{Jackson}:
\begin{equation}
\overline{\braket{\mathcal{H}(\bm{x})}}\equiv\int d^dx'\,F(\bm{x}-\bm{x}')\braket{\mathcal{H}(\bm{x}')}.
\end{equation}
Here, the envelop function $F(\bm{x})$ is normalized to unity,
\begin{equation}
\int d^dx\,F(\bm{x})=1,
\end{equation}
and is localized around the origin.
$F(\bm{x})$ is assumed to vary smoothly in a length scale $\delta l$ much larger than lattice constants.
We also assume $\delta l$ can be taken sufficiently smaller than the length scale we are interested in, that is, $q^{-1}$.
This means that $\phi(\bm{x})$ and its derivatives commute with coarse-graining operations: $\overline{\phi(\bm{x})}=\phi(\bm{x})$, $\overline{\partial_i\phi(\bm{x})}=\partial_i\phi(\bm{x})$, and so on.

Coarse graining of $\braket{\mathcal{H}_0(\bm{x})}$ can be done in the following way.
By using $\bm{x}'=\bm{R}'+\Delta\bm{r}'$,
\begin{subequations}
\begin{align}
\overline{\braket{\mathcal{H}_0(\bm{x})}}&=\sum_{\bm{R}'}\int d^d\!\Delta r'\,F(\bm{x}-\bm{R}'-\Delta\bm{r}')\,\frac{1}{N}\Tr[P(\Delta\bm{r}')Hf(H)]\\
&=\frac{1}{N}\int d^d\!\Delta r'\,\Tr[P(\Delta\bm{r}')Hf(H)]\sum_{\bm{R}'}F(\bm{x}-\bm{R}'-\Delta\bm{r}')\\
&=\frac{1}{N}\int d^d\!\Delta r'\,\Tr[P(\Delta\bm{r}')Hf(H)]\,\frac{1}{\Omega}\int d^d{{R}'}\,F(\bm{x}-\bm{R}'-\Delta\bm{r}')\\
&=\frac{1}{V}\Tr[Hf(H)].
\end{align}
\end{subequations}
Thus, we correctly reproduce average energy density.
From this calculation, we obtain the prescription for coarse graining: We may just replace $P(\Delta\bm{r})$ with the identity operator multiplied with $\Omega^{-1}$.

\subsection{Evaluation of \Eq{eq:p_zx}}
We are now ready to evaluate coefficients Eqs.~\eqref{eq:p_zx}, \eqref{eq:Q_zx}, and \eqref{eq:Qp_zx}.
We first evaluate Eq.~\eqref{eq:p_zx}.
Substituting Fourier transform of Green's function, we obtain
\begin{subequations}
\begin{align}
p_i(z,\bm{x})&=-\int d^d\!\Delta r'\frac{1}{N}\sum_{\bm{k},\bm{k}'}\tr[P(\Delta\bm{r})i\partial_{k_i}G_0(z,\bm{k})\,P(\Delta\bm{r}')G_0(z,\bm{k}')]
\frac{1}{N}\sum_{\bm{R}'}e^{i(\bm{k}-\bm{k}')\cdot(\bm{x}-\bm{x}')}\\
&=-\frac{1}{N}\sum_{\bm{k}}\tr[P(\Delta\bm{r})i\partial_{k_i}G_0(z,\bm{k})\ G_0(z,\bm{k})].
\end{align}
\end{subequations}
After coarse graining, we obtain
\begin{subequations}
\begin{align}
\overline{p_i(z,\bm{x})}&=-\frac{1}{V}\sum_{\bm{k}}\tr[i\partial_{k_i}G_0(z,\bm{k})\ G_0(z,\bm{k})]\\
&=-\frac{1}{2V}\sum_{\bm{k}}i\partial_{k_i}\tr[G_0(z,\bm{k})^2]\\
&=0.
\end{align}
\end{subequations}
We used the fact that $\Tr[G_0(z,\bm{k})^2]$ has the periodicity of the Brilouin zone.
Thus, contribution from $\partial_i\phi(\bm{x})$ to the free energy density identically vanish:
\begin{equation}
\overline{\partial_i\phi(\bm{x})p_i(z,\bm{x})}=\partial_i\phi(\bm{x})\,\overline{p_i(z,\bm{x})}=0.
\end{equation}
\subsection{Evaluation of Eqs.~\eqref{eq:Q_zx} and \eqref{eq:Qp_zx}}
We can evaluate Eqs.~\eqref{eq:Q_zx} and \eqref{eq:Qp_zx} in the same way.
We obtain
\begin{equation}
\overline{Q_{ij}(z,\bm{x})}=\frac{1}{2V}\sum_{\bm{k}}\tr[\partial_{k_i}G_0(z,\bm{k})\ \partial_{k_j}G_0(z,\bm{k})],
\end{equation}
and,
\begin{subequations}
\begin{align}
\overline{Q'_{ij}(z,\bm{x})}&=\frac{1}{2V}\sum_{\bm{k}}\tr[\partial_{k_i}G_0(z,\bm{k})\ G_0(z,\bm{k})\partial_{k_j}G_0(z,\bm{k})+\partial_{k_j}G_0(z,\bm{k})\ G_0(z,\bm{k})\partial_{k_i}G_0(z,\bm{k})]\label{eq:Qp_temp}\\
&=\frac{1}{4V}\sum_{\bm{k}}\tr[\partial_{k_i}G_0(z,\bm{k})^2\ \partial_{k_j}G_0(z,\bm{k})+\partial_{k_j}G_0(z,\bm{k})\ \partial_{k_i}G_0(z,\bm{k})^2]\\
&=-\frac{1}{2}\partial_\mu\overline{Q_{ij}(z,\bm{x})}.
\end{align}
\end{subequations}
In the last line, we used $\partial_\mu G_0(z)=-G_0(z)^2$.
After all, we have only to evaluate the following complex integral, in order to determine thermodynamic electric multipole moments up to EQM:
\begin{equation}
\oint\frac{dz}{2\pi i}zf(z)\,\overline{Q_{ij}(z,\bm{x})}=\frac{1}{2V}\sum_{\bm{k}}\oint\frac{dz}{2\pi i}zf(z)\tr[\partial_{k_i}G_0(z,\bm{k})\ \partial_{k_j}G_0(z,\bm{k})].
\end{equation}
This is easily achieved, and we obtain
\begin{align}
\oint\frac{dz}{2\pi i}zf(z)\,\overline{Q_{ij}(z,\bm{x})}&=\frac{1}{V}\sum_{\bm{k},n}\Bigl[\frac{1}{12}\partial_{k_i}\epsilon_n(\bm{k})\,\partial_{k_j}\epsilon_n(\bm{k})\,\partial_\epsilon^3(\epsilon f(\epsilon))|_{\epsilon\to\epsilon_n(\bm{k})}\notag\\
&\qquad\qquad\qquad+\frac{1}{2}g_n^{ij}(\bm{k})\partial_\epsilon(\epsilon f(\epsilon))|_{\epsilon\to\epsilon_n(\bm{k})}+X_n^{ij}(\bm{k})(\epsilon f(\epsilon))|_{\epsilon\to\epsilon_n(\bm{k})}\Bigr].\label{eq:result_E}
\end{align}

\subsection{From energy density to free energy density}
Now we have obtained explicit expression of the coarse-grained energy density $E(\bm{x})=\overline{\braket{\mathcal{H}(\bm{x})}}$.
In order to obtain the coarse-grained free energy density $F(\bm{x})$, we have to solve the differential equations:
\begin{subequations}
\begin{gather}
\partial_\beta(\beta F(\bm{x}))=E(\bm{x}),\\
\lim_{\beta\to\infty} [F(\bm{x})-E(\bm{x})]=0.
\end{gather}
\end{subequations}
In practice, this is easily solved by the following replacement:
\begin{equation}
E(\bm{x})\to F(\bm{x})\quad \Longleftrightarrow\quad\epsilon f(\epsilon)\to -\int_\epsilon^\infty d\epsilon'\,f(\epsilon')=-\frac{1}{\beta}\log(1+e^{-\beta\epsilon}).
\end{equation}
This replacement is easily done for \Eq{eq:result_E}, and we obtain the results in the main text.

\section{Derivation of EQMs via Kubo formula}
\label{app:kubo}
In this section, we give an alternative derivation of the electric quadrupole moment by using the Kubo formula.
We calculate the energy-density correlation function,
\begin{align}
  \chi_{{\hat H}_0(\mu), {\hat n}}^{\rm R}({\bm q}, \omega)
  = & \sum_{nm} \int \frac{d^d k}{(2 \pi)^d}
  \frac{\epsilon_n({\bm k} - {\bm q}) + \epsilon_m({\bm k})}{2}
  \langle u_n({\bm k} - {\bm q}) | u_m({\bm k}) \rangle \langle u_m({\bm k}) | u_n({\bm k} - {\bm q}) \rangle \notag \\
  & \times \frac{f(\epsilon_n({\bm k} - {\bm q})) - f(\epsilon_m({\bm k}))}{\omega + \epsilon_n({\bm k} - {\bm q}) - \epsilon_m({\bm k}) + i \eta}, \label{eq:kubo1}
\end{align}
which characterizes the response
$\delta \langle {\hat H}_0(\mu) \rangle({\bm q}, \omega) = \chi_{{\hat H}_0(\mu), {\hat n}}^{\rm R}({\bm q}, \omega) \phi({\bm q}, \omega)$.
The auxiliary electric quadrupole moment is identified as
${\tilde Q}^{ij} = \lim_{{\bm q} \rightarrow 0} \lim_{\eta \rightarrow +0} \chi_{{\hat H}_0(\mu), {\hat n}}^{\rm R}({\bm q}, 0)/(i q_i) (i q_j)$.
The intraband contribution for $n = m$ is expanded up to the second order with respect to ${\bm q}$ as
\begin{align}
  \lim_{\eta \rightarrow +0} \chi_{{\hat H}_0(\mu), {\hat n}}^{\rm R (I)}({\bm q}, 0)
  = & \sum_n \int \frac{d^d k}{(2 \pi)^d}
  \frac{\epsilon_n({\bm k} - {\bm q}) + \epsilon_n({\bm k})}{2}
  \langle u_n({\bm k} - {\bm q}) | u_n({\bm k}) \rangle \langle u_n({\bm k}) | u_n({\bm k} - {\bm q}) \rangle
  \frac{f(\epsilon_n({\bm k} - {\bm q})) - f(\epsilon_n({\bm k}))}{\epsilon_n({\bm k} - {\bm q}) - \epsilon_n({\bm k})} \notag \\
  = & \sum_n \int \frac{d^d k}{(2 \pi)^d}
  (\epsilon_n - q_i \partial_{k_i} \epsilon_n/2 + q_i q_j \partial_{k_i} \partial_{k_j} \epsilon_n/4) \notag \\
  & \times (1 - q_i \langle \partial_{k_i} u_n | u_n \rangle + q_i q_j \langle \partial_{k_i} \partial_{k_j} u_n | u_n \rangle/2)
  (1 - q_i \langle u_n | \partial_{k_i} u_n \rangle + q_i q_j \langle u_n | \partial_{k_i} \partial_{k_j} u_n \rangle/2) \notag \\
  & \times \{f^{\prime}(\epsilon_n) - q_i \partial_{k_i} \epsilon_n f^{\prime \prime}(\epsilon_n)/2
  + q_i q_j [3 \partial_{k_i} \partial_{k_j} \epsilon_n f^{\prime \prime}(\epsilon_n) + 2 \partial_{k_i} \epsilon_n \partial_{k_j} \epsilon_n f^{(3)}(\epsilon_n)]/12\} \notag \\
  = & \sum_n \int \frac{d^d k}{(2 \pi)^d}
  \epsilon_n f^{\prime}(\epsilon_n) \notag \\
  & + q_i \sum_n \int \frac{d^d k}{(2 \pi)^d}
  \{\partial_{k_i} \epsilon_n [f^{\prime}(\epsilon_n) + \epsilon_n f^{\prime \prime}(\epsilon_n)]/2
  + [\langle \partial_{k_i} u_n | u_n \rangle + \langle u_n | \partial_{k_i} u_n \rangle] \epsilon_n f^{\prime}(\epsilon_n)\} \notag \\
  & - q_i q_j \sum_n \int \frac{d^d k}{(2 \pi)^d}
  \{[\partial_{k_i} \partial_{k_j} \epsilon_n f^{\prime}(\epsilon_n) + \partial_{k_i} \epsilon_n \partial_{k_j} \epsilon_n f^{\prime \prime}(\epsilon_n)]/4 \notag \\
  & + (\langle \partial_{k_i} u_n | u_n \rangle + \langle u_n | \partial_{k_i} u_n \rangle) \partial_{k_j} \epsilon_n [f^{\prime}(\epsilon_n) + \epsilon_n f^{\prime \prime}(\epsilon_n)]/2 \notag \\
  & + (\langle \partial_{k_i} \partial_{k_j} u_n | u_n \rangle + \langle u_n | \partial_{k_i} \partial_{k_j} u_n \rangle
  + \langle \partial_{k_i} u_n | u_n \rangle \langle u_n | \partial_{k_j} u_n \rangle + \langle \partial_{k_j} u_n | u_n \rangle \langle u_n | \partial_{k_i} u_n \rangle) \epsilon_n f^{\prime}(\epsilon_n)/2\} \notag \\
  & + \epsilon_n [3 \partial_{k_i} \partial_{k_j} \epsilon_n f^{\prime \prime}(\epsilon_n) + 2 \partial_{k_i} \epsilon_n \partial_{k_j} \epsilon_n f^{(3)}(\epsilon_n)]/12\} \notag \\
  = & \sum_n \int \frac{d^d k}{(2 \pi)^d}
  \epsilon_n f^{\prime}(\epsilon_n) \notag \\
  & - q_i q_j \sum_n \int \frac{d^d k}{(2 \pi)^d}
  \{-(\langle \partial_{k_i} u_n | Q_n | \partial_{k_j} u_n \rangle + \langle \partial_{k_i} u_n | Q_n |\partial_{k_j} u_n \rangle) \epsilon_n f^{\prime}(\epsilon_n)/2 \notag \\
  & + \partial_{k_i} \partial_{k_j} \epsilon_n [2 f^{\prime}(\epsilon_n) + \epsilon_n f^{\prime \prime}(\epsilon_n)]/12\}. \label{eq:kubo2}
\end{align}
Here we have dropped the argument ${\bm k}$ of $\epsilon_n({\bm k}), | u_n({\bm k}) \rangle$ for simplicity and introduced $Q_n = 1 - | u_n \rangle \langle u_n |$.
Regarding the first order,
the first term is expressed as the total derivative $\partial_{k_i} [\epsilon_n f^{\prime}(\epsilon_n)]/2$ with respect to $k_i$ and hence can be dropped,
and the second term vanishes thanks to $\langle u_n | u_n \rangle = 1$.
In general, the odd orders vanishes,
because $\lim_{\eta \rightarrow +0} \chi_{{\hat H}_0(\mu), {\hat n}}^{\rm R}({\bm q}, 0)$ is an even function with respect to ${\bm q}$ even when the inversion symmetry is broken.
This fact indicates that the odd-rank electric multipole moments, namely, dipole, octupole, \dots, cannot be defined thermodynamically.

The interband contribution for $n \not= m$ is expanded up to the second order as
\begin{align}
  \lim_{\eta \rightarrow +0} \chi_{{\hat H}_0(\mu), {\hat n}}^{\rm R (II)}({\bm q}, 0)
  = & \sum_{n \not= m} \int \frac{d^d k}{(2 \pi)^d}
  \frac{\epsilon_n({\bm k} - {\bm q}) + \epsilon_m({\bm k})}{2}
  \langle u_n({\bm k} - {\bm q}) | u_m({\bm k}) \rangle \langle u_m({\bm k}) | u_n({\bm k} - {\bm q}) \rangle
  \frac{f(\epsilon_n({\bm k} - {\bm q})) - f(\epsilon_m({\bm k}))}{\epsilon_n({\bm k} - {\bm q}) - \epsilon_m({\bm k})} \notag \\
  = & q_i q_j \sum_{n \not= m} \int \frac{d^d k}{(2 \pi)^d}
  \frac{\epsilon_n + \epsilon_m}{2}
  \langle \partial_{k_i} u_n | u_m \rangle \langle u_m | \partial_{k_j} u_n \rangle
  \frac{f(\epsilon_n) - f(\epsilon_m)}{\epsilon_n - \epsilon_m} \notag \\
  = & q_i q_j \sum_{n \not= m} \int \frac{d^d k}{(2 \pi)^d}
  \frac{\epsilon_n + \epsilon_m}{2}
  (\langle \partial_{k_i} u_n | u_m \rangle \langle u_m | \partial_{k_j} u_n \rangle + \langle \partial_{k_j} u_n | u_m \rangle \langle u_m | \partial_{k_i} u_n \rangle)
  \frac{f(\epsilon_n)}{\epsilon_n - \epsilon_m}. \label{eq:kubo3}
\end{align}
From Eqs.~\eqref{eq:kubo2} and \eqref{eq:kubo3}, we obtain
\begin{equation}
  {\tilde Q}^{ij}
  = \sum_n \int \frac{d^d k}{(2 \pi)^d} \{X_n^{ij} \epsilon_n f(\epsilon_n) + g_n^{ij} [f(\epsilon_n) + \epsilon_n f^{\prime}(\epsilon_n)]/2
  - (m^{\ast -1}_n)^{ij} [2 f^{\prime}(\epsilon_n) + \epsilon_n f^{\prime \prime}(\epsilon_n)]/12\}. \label{eq:kubo4}
\end{equation}
$X_n^{ij}$, the quantum metric $g_n^{ij}$, and the inverse effective mass $(m^{\ast -1}_n)^{ij}$ are defined in the main text.
The thermodynamic electric quadrupole moment, which is defined by the grand potential, is expressed as
\begin{equation}
  Q^{ij}
  = \sum_n \int \frac{d^d k}{(2 \pi)^d} \left[-X_n^{ij} \int_{\epsilon_n}^{\infty} {\rm d} \epsilon f(\epsilon) + \frac{1}{2} g_n^{ij} f(\epsilon_n)
  - \frac{1}{12} (m^{\ast -1}_n)^{ij} f^{\prime}(\epsilon_n)\right]. \label{eq:kubo5}
\end{equation}

\section{Calculation of particle density and thermodynamic relation}
\label{app:ThermoRel}
In order to see the relation of EQM with charge density, we can use the following formula:
\begin{equation}
  \rho(\bm{x})=-\frac{\delta}{\delta \phi(\bm{x})}\int\,d\bm{x}'F(\bm{x}').
\end{equation}
We obtain from \Eq{eq:eq4},
\begin{align}
  \rho(\bm{x})&=\rho_0(\xi-\phi(\bm{x}))+\partial_i\partial_jQ^{ij}(\xi-\phi(\bm{x}))-\frac{1}{2}\frac{\partial^2 Q^{ij}(\xi-\phi(\bm{x}))}{\partial^2\xi}\partial_i\phi(\bm{x})\partial_j\phi(\bm{x})\\
  &=\rho_0(\xi-\phi(\bm{x}))+\partial_i\partial_jQ^{ij}(\xi-\phi(\bm{x}))+O(\nabla\phi(\bm{x}))^2.
\end{align}
This is also reproduced by calculating $-\partial_\xi F(\bm{x})$.
With electric field $E_i(\bm{x})=-\partial_i\phi(\bm{x})$, this is rewritten as
\begin{equation}
  \rho(\bm{x})-\rho_0(\xi)=-\partial_i\left[-\frac{\partial Q_{ij}(\xi-\phi(\bm{x}))}{\partial\xi}E_j(\bm{x})\right].\label{eq:chargedensity}
\end{equation}
This gives the polarization charge in local equilibrium under local modulation of the scalar potential.

In insulators at zero temperature, insertion of local electric field can be done adiabatically.
Thus, the above relation should also be recovered with the increase of charge density under physical process of electric-field insertion, which is calculated with Kubo formula.
This leads to the identity
\begin{equation}
  -\left.\frac{\partial Q_{ij}(\xi)}{\partial\xi}\right|_{T\to0,\,\mathrm{insulator}}=\chi_{ij}^e.
\end{equation}
Note also that $\partial^2 Q^{ij}(\xi)/\partial\xi^2$ vanishes for this situation.

\section{Gauge invariance of the formulas for EQMs}
\label{app:app3}
Here, we make some comments on the gauge invariance of \Eq{eq:Result2}.
In \Eq{eq:Result2}, gauge invariance is manifest when all the bands are isolated with each other, since the interband Berry connection $A_{nm}^i(\bm{k})$ is gauge invariant.
The question is whether gauge invariance is preserved when the system has degeneracy at some $\bm{k}$ point $\bm{k}_0$.
Actually, we have to be careful to interpret \Eq{eq:Result2}, since $g_n^{ij}(\bm{k})$ and $X_n^{ij}(\bm{k})$ are gauge-dependent at $\bm{k}_0$, and furthermore, $X_n^{ij}(\bm{k}_0)$ is divergent.

In order to answer this question, we discuss the behavior of the $\bm{k}$ integrand when the band gap $\Delta\epsilon_{nm}(\bm{k})=\epsilon_n(\bm{k})-\epsilon_m(\bm{k})$ is small.
It is convenient to use an alternative expression of \Eq{eq:Result2}:
  \begin{subequations}
  \begin{gather}
Q_{ij}(\mu)=\frac{1}{2}\int_{\mathrm{BZ}}\frac{d^dk}{(2\pi)^d}\sum_{m n}(V_{nm}^i(\bm{k})V_{mn}^j(\bm{k})+c.c.)\,\mathcal{Q}_{nm}(\bm{k}),\label{eq:Result3a}\\
\mathcal{Q}_{nm}(\bm{k})=
\frac{1-\delta_{nm}}{(\epsilon_n(\bm{k})-\epsilon_m(\bm{k}))^2}\left(\frac{f(\epsilon_n(\bm{k}))+f(\epsilon_m(\bm{k}))}{2}-\frac{1}{\epsilon_n(\bm{k})-\epsilon_m(\bm{k})}\int_{\epsilon_m(\bm{k})}^{\epsilon_n(\bm{k})}d\epsilon\,f(\epsilon)\right)
+\frac{\delta_{nm}}{12}f''(\epsilon_m(\bm{k})).\label{eq:Result3b}
\end{gather}
\label{eq:Result3}
\end{subequations}
Here, $V_{nm}^i(\bm{k})=\braket{u_n(\bm{k})|\partial_{k_i}H(\bm{k})|u_m(\bm{k})}.$
By expanding $\mathcal{Q}_{nm}(\bm{k})$ in terms of $\Delta\epsilon_{nm}(\bm{k})$, we obtain
\begin{equation}
  \mathcal{Q}_{mn}(\bm{k})=\frac{1}{12}f''(\epsilon_m(\bm{k}))+\frac{1-\delta_{nm}}{24}f'''(\epsilon_m(\bm{k}))\Delta\epsilon_{nm}(\bm{k})+\frac{1-\delta_{nm}}{80}f''''(\epsilon_m(\bm{k}))\Delta\epsilon_{nm}(\bm{k})^2,
  \label{eq:Qnm}
\end{equation}
up to $O(\Delta\epsilon_{nm}(\bm{k})^2)$.
Thus, $\lim_{\bm{k}\to\bm{k}_0}\mathcal{Q}_{nm}(\bm{k})=f''(\epsilon_m(\bm{k}_0))/12$ holds.
By using this, gauge-invariance as well as non-divergent behavior of the $k$-integrand at $\bm{k}=\bm{k}_0$ is readily shown by taking the summation over $n$ and $m$.
Thus, we have established gauge invariance and regularity of $Q_{ij}(\mu)$.
We note that Eqs.~\eqref{eq:Result3} and \eqref{eq:Qnm} are useful for numerical calculations, since they are manifestly gauge invariant.

It should be noticed that Eqs.~\eqref{eq:Result3} and \eqref{eq:Qnm} are valid also when the system has perfect degeneracy on the entire Brillouin zone.
For example, let us consider a system with inversion and time-reversal symmetries.
In principle, we can apply magnetic field to lift all the degeneracy (if necessary, also by applying other perturbation).
Then, we may use Eqs.~\eqref{eq:Result3} and \eqref{eq:Qnm} to evaluate EQM.
By taking the limit of $\Delta\epsilon_{mn}(\bm{k})=0$, we obtain the contribution from a pair of the Kramers-degenerate bands $D$ as
\begin{align}
  &\sum_{nm\in D}V_{nm}^i(\bm{k})V_{mn}^j(\bm{k}_0)f''(\epsilon_m(\bm{k}))\\
  &=f''(\epsilon_D(\bm{k}))\tr[P_D(\bm{k})\partial_{k_i}H(\bm{k})P_D(\bm{k})\partial_{k_j}H(\bm{k})],\label{eq:temp17}
\end{align}
for all $\bm{k}$.
Here, we denoted the energy dispersion as $\epsilon_D(\bm{k})(=\epsilon_m(\bm{k})=\epsilon_n(\bm{k}))$, and the projection operator onto the bands $D$ as $P_D(\bm{k})$.
By using $H(\bm{k})=\epsilon_D(\bm{k})P_D(\bm{k})+Q_D(\bm{k})H(\bm{k})Q_D(\bm{k})$ with $Q_D(\bm{k})=1-P_D(\bm{k})$,
\Eq{eq:temp17} is simplified to $f''(\epsilon_D(\bm{k}))\partial_{k_i}\epsilon_D(\bm{k})\partial_{k_j}\epsilon_D(\bm{k})$.
Thus, for the case of Kramers degeneracy (and for perfect degeneracy in general), we can simply drop the Kramers partner from the summation range of $m$ in Eqs.~\eqref{eq:quantum_metric} and \eqref{eq:Xnij} to evaluate \Eq{eq:Result2}, as well as in the first term of \Eq{eq:Result3b} to evaluate \Eq{eq:Result3a}.

\section{Electric susceptibility and Lindhard function} \label{app:lindhard}
The third term of \Eq{eq:Result2} originates from a generalization of the Lindhard function,
which is the density-density correlation function for a simple Hamiltonian $\hat{H} =\hat{p}^2/2 m$.
In three dimensions, the Lindhard function for $\omega = 0$ is expressed as
\begin{equation}
  F(q)
  = -\frac{n_{\rm F}}{\epsilon_{\rm F}} \frac{3}{4} \left[1 + \frac{1 - (q/2 k_{\rm F})^2}{q/k_{\rm F}} \ln \left|\frac{1 + q/2 k_{\rm F}}{1 - q/2 k_{\rm F}}\right|\right]
  = -\frac{3 n_{\rm F}}{2 \epsilon_{\rm F}} [1 - (q/k_{\rm F})^2/12] + o(q^2). \label{eq:lindhard2}
\end{equation}
Here, $k_{\rm F}$ is the Fermi wave number, $\epsilon_{\rm F} = k_{\rm F}^2/2 m$ is the Fermi energy,
and $n_{\rm F} = k_{\rm F}^3/6 \pi^2$ is the electron density without the spin degree of freedom.
Since ${\rm d} n_{\rm F}/{\rm d} \epsilon_{\rm F} = 3 n_{\rm F}/2 \epsilon_{\rm F}, {\rm d}^2 n_{\rm F}/{\rm d} \epsilon_{\rm F}^2 = 3 n_{\rm F}/4 \epsilon_{\rm F}^2$,
we obtain
\begin{equation}
  F(q)
  = -\frac{{\rm d} n_{\rm F}}{{\rm d} \epsilon_{\rm F}} + \frac{{\rm d}^2 n_{\rm F}}{{\rm d} \epsilon_{\rm F}^2} \frac{q^2}{12 m} + o(q^2). \label{eq:lindhard3}
\end{equation}
The coefficient of $q^2$ is equal to $\mu$ derivative of the third term in Eq.~\eqref{eq:Result2} at zero temperature.
\end{widetext}

\end{document}